\documentclass[epj]{svjour} 

\usepackage{graphicx}
\usepackage{float}
\usepackage{amsmath}
\usepackage{amssymb}
\usepackage{mathtools}
\usepackage{amsfonts}
\usepackage{dsfont}
\usepackage{array}
\usepackage{bm}
\usepackage{mathrsfs}
\usepackage{pifont}
\usepackage{multirow}
\usepackage{upgreek}
\usepackage[dvipsnames]{xcolor}
\usepackage[pdftitle={On the equivalence of Prony and Lanczos methods for
    Euclidean correlation functions}
  pdfauthor={Johann Ostmeyer and Aniket Sen and Carsten Urbach},
  bookmarks,
  colorlinks,
  linkcolor=myblue,
  citecolor=mymagenta,
  menucolor=black,
  urlcolor=myblue,
  plainpages=false,
  pdfpagelabels,
  hypertexnames=false]{hyperref}
\usepackage{verbatim}
\usepackage{slashed}
\usepackage{algorithm}
\usepackage{algorithmicx}
\usepackage{algpseudocode}

\usepackage{subfigure}

\usepackage[backend=bibtex,terseinits=false,sorting=none,style=phys,biblabel=brackets,
  defernumbers=true,minnames=15,maxnames=99,
  minsortnames=15, maxsortnames=99,
  minbibnames=15, maxbibnames=99]{biblatex}
\definecolor{mymagenta}{RGB}{200, 0, 100}
\definecolor{myblue}{RGB}{45, 48, 146}

\graphicspath{{plots/}}

\begin{document}
\title{On the equivalence of Prony and Lanczos methods for Euclidean
  correlation functions}

\author{J.~Ostmeyer\inst{1}
  \and A.~Sen\inst{1}
  \and C.~Urbach\inst{1}
}
\authorrunning{J.~Ostmeyer \emph{et al.}}
\titlerunning{On the equivalence of Prony and Lanczos \ldots}
\institute{
  Helmholtz-Institut~für~Strahlen-~und~Kernphysik~and~Bethe~Center~for
  Theoretical~Physics, Universität~Bonn, \\ Bonn,~Germany
}

\date{\today}
\abstract{
  We investigate the oblique Lanczos method recently put forward in
  Ref.~\cite{Wagman:2024rid} for analysing Euclidean correlators in
  lattice field theories and show that it is analytically equivalent
  to the well known Prony Generalised Eigenvalue Method
  (PGEVM). Moreover, we discuss that the signal-to-noise problem is
  not alleviated by either of these two methods. Still, both methods
  show clear advantages when compared to the standard effective mass
  approach. 
   }

\maketitle

\section{Introduction}

In lattice field theory, Euclidean two-point correlation functions $C$ 
represent the central objects to be determined numerically in
Monte Carlo simulations. They are estimated from vacuum expectation
values of local operators $O_i, O_j$ with appropriate quantum numbers
\begin{equation}
  \label{eq:correlator}
  C_{ij}(t-t^\prime)\ =\ \langle\, O_i^\dagger(t)\ O_j(t^\prime)\, \rangle\,.
\end{equation}
They allow one to access energy eigenvalues of the lattice Hamiltonian and
operator matrix elements via the spectral decomposition for energy
levels $0< E_1<E_2<\ldots$ for instance for a single operator $O$
\begin{equation}
  \label{eq:spectral}
  C(t)\ =\ \sum_n |\langle 0|O|n\rangle|^2 e^{-E_n
  t}\,.
\end{equation}
The ground state energy level can then be determined from $C(t)$ at
sufficiently large $t$.

Since typically $C(t)$ is being determined in stochastic Monte Carlo
simulations, also the statistical uncertainty needs to be estimated
from the corresponding variance. As discussed by Lepage in
Ref.~\cite{Lepage:1989}, the variance can be understood as a
correlation function itself decaying exponentially with a potentially
smaller energy level $E_1^\prime$, leading to the infamous
signal-to-noise (StN) problem
\begin{equation}
  \operatorname{StN}(t)\ \propto\ e^{-\Delta E\, t},
\end{equation}
with $\Delta E = E_1 - E_1^\prime/2$. The most famous example is
probably the nucleon two-point function, which decays at large $t$
proportional to $\exp(-M_N t)$, and the variance with $\exp(-3M_\pi
t)$. Therefore, the StN ratio for the nucleon is decreasing
exponentially with $\exp(-(M_N -3M_\pi/2)t)$. The most famous exception
from this StN problem is the pion two-point function, which can be
shown to have a constant StN ratio at large $t$-values.

This StN problem triggered a lot of effort to find novel methods which
either allow to determine energy levels at earlier
Euclidean times, or to tame the increasing StN ratio. One widely used
method in the former class is the Generalised Eigenvalue Method
(GEVM)~\cite{Michael:1982gb,Luscher:1990ck,Blossier:2009kd}, which
however requires a correlator matrix  
$C_{ij}(t)$. The Prony GEVM (PGEVM)~\cite{Fischer:2020bgv} (see
Ref.~\cite{Prony:1795} for the original paper) or
generalised pencil of function method, on the other hand, can be used
with single correlation functions and allows to work at earlier times,
too. For a method based on ordinary differential equations see
Ref.~\cite{Romiti:2019qim}. 

Recently, the author of Ref.~\cite{Wagman:2024rid} applied the
Lanczos method to this problem. Wagman argues in the paper that the
Lanczos method does both, allow extraction of energy levels at early
$t$-values and to tame the StN problem at late $t$-values.

In this paper we show that Lanczos is equivalent to the PGEVM, also on
noisy data. Moreover, we cannot confirm that the StN problem is
alleviated by either of the methods. Still, both Lanczos and PGEVM
appear to be very useful in controlling systematic effects.

\section{Methods}

The simplest way to analyse Euclidean correlation functions is via the
$\log$-effective mass
\begin{equation}
  M_\mathrm{eff}(t)\ =\ -\log\left(\frac{C(t+1)}{C(t)}\right)\,,
  \label{eq:Meff}
\end{equation}
applicable to correlation functions without back-propagating states.
For correlators periodic in time taking the $\log$ is not sufficient,
which is why one numerically inverts $f_t(M) =
\cosh(M(t+1))/\cosh(Mt)$ for $M$ to define
\begin{equation}
  \label{eq:MeffInv}
  \hat{M}_\mathrm{eff}(t) = [f_t(M)]^{-1}\,.
\end{equation}

\subsection{Oblique Lanczos}

The oblique Lanczos method is well known and was applied to
correlation functions $C(t)$ Eq.~\ref{eq:correlator} for the first
time by Wagman in Ref.~\cite{Wagman:2024rid}.  The algorithm we have
implemented is summarised in listing~\ref{alg:lanczos}. To our
understanding this is the algorithm the author has used to obtain the
results presented in Ref.~\cite{Wagman:2024rid}. Note that we refer here to the
original version of Ref.~\cite{Wagman:2024rid} posted to the arXiv.
For the application of Lanczos to the estimation of matrix elements
see Ref.~\cite{Hackett:2024xnx}.

The Lanczos method produces sets of bi-orthogonal vectors
$|v_i\rangle$ and $|w_j\rangle$ which are used to construct
at iteration step $n$ the tri-diagonal matrix
\begin{equation}
  \label{eq:Tn}
  T_n\ =\ W_n^t T V_n^{~}\,.
\end{equation}
Note that in contrast to Ref.~\cite{Wagman:2024rid} we use $n$ as the
iteration count instead of $m$. The elements of $T_n$ can be
computed from $2n$ elements of the correlation function $C(0), C(1),
C(2), \ldots$
without the need to explicitly construct the $|v_i\rangle$ and
$|w_j\rangle$. 

The eigenvalues $\{\lambda_i\}$ of $T_n$ are equal to the eigenvalues of $T$ if the
rank of $T$ is $n$. For rank of $T$ larger than $n$ the eigenvalues of
$T_n$ approximate the eigenvalues of $T$. For details on the
convergence see Ref.~\cite{Wagman:2024rid} or the mathematical
literature. The eigenvalues $\lambda_i$ are related to the ground
state energy via
\[
E_n = -\log(\lambda_\mathrm{max})
\]
with $\lambda_\mathrm{max}$ the largest real eigenvalue smaller than 1.

\begin{algorithm}[t]
  \caption{Oblique Lanczos for LFT}
  \label{alg:lanczos}
  \begin{algorithmic}[1]
    \State \textbf{Input:} Correlator $C(t)$ for $t=0, ..., 2N-1$.
    \For{$n=1,\ldots, N$}
    \State set $A_1^k = C(k)/C(0)\,,\quad k>0$
    \State $\alpha_1 = A_1^1\,,\ \beta_1=\gamma_1=0$
    \State $B_1^k = G_1^k = 0$
    \For{$j=1,\ldots, n-1$}
    \State use eq.~(53) Ref.~\cite{Wagman:2024rid}
    \[
    \langle s_{j+1}|r_{j+1}\rangle = A_j^2
    -\alpha_j^2-\beta_j\gamma_j
    \]
    \State use eq~(41) Ref.~\cite{Wagman:2024rid}
    \[
    \rho_{j+1} = \sqrt{|\langle
      s_{j+1}|r_{j+1}\rangle|}\,,\quad\tau_{j+1} = \frac{\langle
      s_{j+1}|r_{j+1}\rangle}{\rho_{j+1}}
    \]
    \State Set $k_\mathrm{max} = 2(n-j)+1$
    \For{$k=2,\ldots,k_\mathrm{max}$}
    \State Use eqs.(44) and (45) Ref.~\cite{Wagman:2024rid}
    \[
    \begin{split}
        G_{j+1}^k &= \frac{1}{\tau_{j+1}}\left(A_j^{k+1} - \alpha_j A_j^k -  \gamma_j B_j^k\right)\\
        B_{j+1}^k &= \frac{1}{\rho_{j+1}}\left(A_j^{k+1} -\alpha_j A_j^k - \beta_jG_j^k \right)\\
    \end{split}
    \]
    \EndFor
    \For{$k=2,\ldots,k_\mathrm{max}$}
    \State Use eq.(46) Ref.~\cite{Wagman:2024rid} and note that $A_0^k$ is not needed, because $\gamma_1=\beta_1=0$
    \[
    \begin{split}
      A_{j+1}^k =& \frac{1}{\tau_{j+1}\rho_{j+1}}\left( A_j^{k+2} -
      2\alpha_j A_j^{k+1}\right.\\
      &+ \alpha_j^2 A_j^k + \alpha_j(\beta_jG_j^k + \gamma_jB_j^k)\\
      &\left.- (\beta_jG_j^{k+1}+ \gamma_j G_j^{k+1}) +
      \gamma_j\beta_jA_{j-1}^k\right)\\
    \end{split}
    \]
    \EndFor
    \State Set
    \[
      \alpha_{j+1} = A_{j+1}^1\,,\gamma_{j+1}=G_{j+1}^1\,,\beta_{j+1}=B_{j+1}^1
    \]
    \EndFor
    \State Diagonalise the tri-diagonal matrix eq.(48) Ref.~\cite{Wagman:2024rid} and
    obtain eigenvalues $\{\lambda_i\}$
    \State Reduce to the set $\{\lambda_i:\,0<\lambda_i<1\,,\ \operatorname{Im}(\lambda_i)=0\}$
    
    \State Set $E_n = -\log(\lambda_\mathrm{max})$ 
    \EndFor
  \end{algorithmic}
\end{algorithm}

\subsection{Prony GEVM}

The second method we consider is the PGEVM~\cite{Fischer:2020bgv},
which is a variant of the Prony method~\cite{Prony:1795}, see also
Refs.~\cite{Fleming:2004hs,Beane:2009kya,Cushman:2019hfh,Cushman:2019tcv,Fleming:2023zml}. It is
equivalent to the so-called generalised pencil of function method
(GPOF). For
matrix elements there is a discussion in Ref.~\cite{Ottnad:2017mzd}.
For the PGEVM first a $n\times n$ symmetric Hankel matrix
\begin{equation}
  H_{ij}(t) = C(t + i\Delta + j\Delta)\,,\ 0\leq i,j < n\,,
  \label{eq:hankel}
\end{equation}
is constructed from the correlation function $C(t)$ for each $t$ with
$\Delta>0$ an additional parameter. Note that $H$ is symmetric, but
for noisy data not necessarily positive definite.
Next, the following GEVP
\begin{equation}
  H(t+\delta t)\cdot v_l(t)\ =\ \Lambda_l^n(t, \delta t) H(t)\cdot v_l(t)
\end{equation}
is solved for eigenvalues $\Lambda_l^n$ and eigenvectors $v_l$, with
another constant parameter $\delta t\geq 1$. The $\Lambda_l^n$ can be
shown to have the form 
\begin{equation}
  \Lambda_l^n(t, \delta t)\ =\ e^{-E_l \delta t}
\end{equation}
independent of $t$ if $n$ is large enough to resolve all relevant
states contributing to $C(t)$, see Ref.~\cite{Fischer:2020bgv} and
references therein. For convenience, we then define
\begin{equation}
  \tilde{M}_n(t)\ =\ -\frac{\log\left(\Lambda_0^n(t,\delta
    t)\right)}{\delta t}\,.
\end{equation}
In practice one often chooses $\delta t$ odd and $\Delta$ even,
such that $H(t)$ and $H(t+\delta t)$ contain disjoint elements of
$C(t)$. 

The method analogous to Lanczos discussed above is to change 
$n=1, \ldots, n_\mathrm{max}$ for $\delta t$ and $\Delta$
fixed. The corresponding algorithm we implemented is
summarised in listing~\ref{alg:pgevm}.

\begin{algorithm}[t]
  \caption{PGEVM for LFT}
  \label{alg:pgevm}
  \begin{algorithmic}[1]
    \State \textbf{Input:} $t_0$, $\delta t$, $n_\text{max}$, and Correlator
    $C(t)$ for $t=0, ..., T-1$
    \For{$n=1, \ldots, n_\mathrm{max}$}
    \State construct $H(t_0+\delta t)$ and $H(t_0)$
    \State determine $\{\Lambda_l^n\}$ from
    \[
    H(t_0+\delta t)\cdot v_l(t_0)\ =\ \Lambda_l^n(t_0, \delta t) H(t_0)\cdot v_l(t_0)
    \]
    \State Reduce to the set
    $\{\Lambda_l^n:\,0<\Lambda_l^n<1\,,\ \operatorname{Im}(\Lambda_l^n)=0\}$
    \State Set $\tilde{M}_n(t) =
    -\log(\Lambda_\mathrm{max}^n)/\delta t$
    \EndFor
  \end{algorithmic}
\end{algorithm}

\subsection{Relation between Lanczos and PGEVM}

Let us first consider the PGEVM. The Hankel matrix $H(t)$
Eq.~\ref{eq:hankel} 
of size $n\times n$ contains the $2n-1$ correlator elements $C(t),
C(t+\Delta), \ldots, C(t+2(n-1)\Delta)$. An identical number of
correlator elements enters $H(t+\delta t)$. However, depending on the
values of $\Delta$ and $\delta t$, there can be significant
overlap in the correlator elements entering $H(t)$ and $H(t+\delta
t)$.

If one chooses $\Delta =1$ and $\delta t = 1$, there are $2n$
correlator elements entering in total, namely $C(t_0), C(t_0+1), \ldots,
C(t_0+2n-1)$. Thus, choosing $t_0=0$, $\Delta=1$ and $\delta t=1$ the PGEVM
is based on the same correlator input like the oblique Lanczos for
the same $n$. In fact, PGEVM and Lanczos are exactly equivalent and,
therefore, expected to yield identical results. The argument is as
follows: following the notation of Ref.~\cite{Wagman:2024rid}, the
correlator Eq.~\ref{eq:correlator} for $O_i=O_j=O$ reads
\begin{equation}
  C(i) = \langle\psi| T^i |\psi\rangle\,,
\end{equation}
with $T=\exp(-\mathcal{H}a)$ the Euclidean time evolution operator,
$a$ the lattice spacing, and
$\mathcal{H}$ the lattice Hamiltonian. Now, both methods compute the
projection of $T$ on the space spanned by the vectors
\[
|\varphi_0\rangle,\,|\varphi_1\rangle\,\ldots,|\varphi_{n-1}\rangle\,,
\]
with
\[
|\varphi_i\rangle\ =\ T^i|\psi\rangle\,.
\]
Let $P_n$ be the column matrix of the $n$ vectors $|\varphi_i\rangle,
i=0,\ldots,n-1$. Then 
\begin{equation}
  H(0) = P_n^t\cdot P_n^{~}\,,\quad H(1) = P_n^t\cdot T\cdot P_n^{~}\,.
\end{equation}
This is possible because
\[
\langle\varphi_0|\varphi_i\rangle = \langle\varphi_1|\varphi_{i-1}\rangle =
\ldots = \langle \varphi_i|\varphi_0\rangle\,,
\]
which is why one can use $2n-1$ correlator elements to compute $H(0)$
(and likewise $H(1)$), without the need to explicitly compute $P$.

Let $|\chi_i\rangle$ be an orthonormal basis of the space spanned by
the $|\varphi_i\rangle$, $T_\chi$ the projection of $T$ to this basis,
and $P_{n,\chi}^t = P_n^t\cdot \chi_n$ with $\chi_n$ the column matrix
of the $|\chi_i\rangle$.
Then
\begin{equation}
  H(0)^{-1}\cdot H(1) = P_{n,\chi}^{-1}\cdot T_\chi\cdot P_{n,\chi}^{~}\,.
\end{equation}
Therefore, $H(0)^{-1}\cdot H(1)$ is similar to $T$ projected to the
space spanned by the $|\varphi_i\rangle$. 
Symmetric Lanczos, on the other hand, computes the tri-diagonal matrix 
\[
T_n\ =\ V_n^t T V_n^{~}\,,
\]
which is again similar to $T$ projected to the very same space. The 
matrix $V_n$ with $V_n^t\cdot V_n^{~}=1$ is constructed as a column
matrix of the vectors $|v_j\rangle$ which are defined via the Lanczos
recursion in terms of the vectors $|\varphi_i\rangle$.

For the case the correlator is given by $C(t) =
\langle\xi|T^t|\psi\rangle$, analogous arguments lead to
\[
  H(0) = Q_n^t\cdot P_n^{~}\,,\quad H(1) = Q_n^t\cdot T\cdot P_n^{~}\,.
\]
It follows like before that $H(0)^{-1}\cdot H(1)=P_n^{-1} T P_n^{~}$ is
similar to $T$ projected to the corresponding subspace.

In Ref.~\cite{Fischer:2020bgv} the corrections to $\Lambda_l^n$ have
been discussed, which stem from not resolved states. For $t_0 >
(t_0+\delta t)/2$ these corrections are exponential as $\exp(-\Delta
E_{m,l}(t_0+\delta t))$ with
\[
\Delta E_{m,l} = E_m - E_l\,.
\]
and $E_m$ the energy of the first unresolved state. With $t_0=0$ and
$\delta t=1$ as we are going to use here, one is formally not in this
regime, and even the expansion used to arrive at this result is not
applicable: convergence is observed in $n\to\infty$ not in
$t_0\to\infty$ or $\delta t\to\infty$.
Instead, the well-known Kaniel-Paige-Saad bound~\cite{Kaniel1966,Paige1971TheCO,Saad1980}
can be used, see Ref.~\cite{Wagman:2024rid}, to show that the deviation
of a given energy $E_l$ scales at most as
\begin{align*}
	\exp\left(-4\sqrt{\Delta E_{l+1,l}\delta t}\,n\right)\,.
\end{align*}
Due to the equivalence of Lanczos and PGEVM, both methods have 
the same convergence properties.

\subsection{Selection of Eigenvalues}

For the statistical analysis we apply the bootstrap, 
thus, for $R$ bootstrap samples we compute
\begin{equation}
  M_\mathrm{eff}^{\star,b}(t)\,,\ E_n^{\star,b}\,,\ \tilde{M}_\mathrm{eff}^{\star,b}(t)\,,
\end{equation}
for $r=1, \ldots, R$, and all $n$-values. The standard
error is then estimated by the standard deviation of the bootstrap 
distribution
\begin{equation}
  \mathrm{err}^\star(O)\ =\ \mathrm{sd}_R(O^{\star,r})
\end{equation}
for $O^{\star,r}$ one of the observables from above. The bootstrap
bias for observable $O$ is defined as
\begin{equation}
  \mathrm{bias}^\star(O)\ =\ \langle O\rangle - \langle O\rangle^\star\,,
\end{equation}
where $\langle O\rangle^\star$ is the bootstrap estimate and $\langle
O\rangle$ the standard estimate of $O$. The bootstrap
estimate of bias can be used to correct for a bias, which means to use
the bootstrap estimator instead of the standard estimator.
In particular, it is well known that the mean is not a
stable estimator of the expectation value for distributions with
outliers, for which the median $\mu$ can be used instead. While the
$\mu$ is not easily estimated on the original data, its bootstrap
estimate is easily computable. However, estimating the error of the
median requires the double bootstrap~\footnote{We acknowledge a very
useful discussion with Daniel Hackett, which only led us to implement the
double bootstrap despite initial reluctance. For a first application
of double bootstrap to Lanczos see the third arXiv version of
Ref.~\cite{Wagman:2024rid}}. In short, each 
bootstrap  replica is resampled again $R_d$ times, leading to double
bootstrap estimates $O^{\star\star r r_d}$, which can be used to
compute the median $\mu^{\star\star r}$ over the $R_d$ double
bootstrap samples for each $r$. Then, the uncertainty of the median is
given by
\[
\mathrm{err}^{\star\star}(\mu)\ =\
\mathrm{sd}_R(\mu^{\star\star r}) \,.
\]
For a brilliant text book on the bootstrap see Ref.~\cite{efron:1994}.
The reason for discussing this in some detail is that noise in the
correlator leads to the fact that the empirical Euclidean transfer
matrix $T_e$ can have negative and even complex eigenvalues. Such
eigenvalues are not physical. They need to be removed in practical
applications, see line 19 of listing~\ref{alg:lanczos} and line 5 of
listing~\ref{alg:pgevm}. Even worse, noise might lead to additional
unphysical real and positive eigenvalues, or also missing ones. For
the identification of these spurious eigenvalues there are important
bounds and sophisticated strategies discussed in
Ref.~\cite{Wagman:2024rid}.

It turns out that bias removal, or using the median as estimator for
the expectation value, is essential for applying both Lanczos and PGEVM
in the way described above. In fact, we find that using the double
bootstrap procedure is most stable for both Lanczos and PGEVM.

However, for comparison reasons to Ref.~\cite{Wagman:2024rid} we also
apply here two more data driven methods to deal with outliers in the
bootstrap distributions: 
\begin{itemize}
\item \emph{outlier removal}: we perform a standard outlier removal
  procedure on the empirical bootstrap distribution according to
  \begin{equation}
    (Q^\star_{25}-1.5\cdot\operatorname{iqr}) < \lambda <  (Q^\star_{75} +
    1.5\cdot\operatorname{iqr}^\star)\,, 
  \end{equation}
  with $Q_{25}^\star$ the $0.25$-quantile and $Q_{75}^\star$ the
  $0.75$-quantile of the bootstrap
  distribution. $\operatorname{iqr}^\star$ represents the
  interquartile range of the bootstrap distribution.

  This procedure has the disadvantage that the number of bootstrap
  samples is changed.

\item \emph{confidence interval}: use the $0.32$- and $0.68$-quantiles
  of the empirical bootstrap distribution to estimate the uncertainty
  instead of the standard estimator $\mathrm{err}^\star$.

\end{itemize}
Finally, as also used in Ref.~\cite{Wagman:2024rid}, the eigenvalue
selection can be guided by specifying a pivot value $L_p$, and chosing
the eigenvalue closest to $L_p$ instead of the maximal eigenvalue
smaller than $1$. The author of Ref.~\cite{Wagman:2024rid} implemented
this to our understanding such that the result obtained on the
original data is used as a pivot element for the bootstrap analysis,
and so we do here for Lanczos, too.\\
For PGEVM we have implemented it as follows: we compute all
real eigenvalues in the range $[0,1]$ on the original data and on all the
bootstrap samples. Then we compute $L_p$ as the median over those
eigenvalues closest to one. Only thereafter we chose on each sample
and the original data the eigenvalue closest to $L_p$ separately.

In the following we will denote the oblique Lanczos with
confidence interval for the error estimate but without pivot element
as \emph{Lanczos a}, with outlier-removal and pivot element as
\emph{Lanczos b}, and with confidence interval and pivot element as
\emph{Lanczos c}. All three methods use bias correction unless
specified otherwise. 

All those three methods as well as the double bootstrap and the
corresponding statistical analyses are implemented in the
\emph{hadron} R-package~\cite{hadron:2024}, which 
is available as open source software.

\section{Numerical Experiments}

\subsection{Synthetic Data}

\begin{figure*}
  \centering
  \includegraphics[width=0.45\textwidth, page=1]{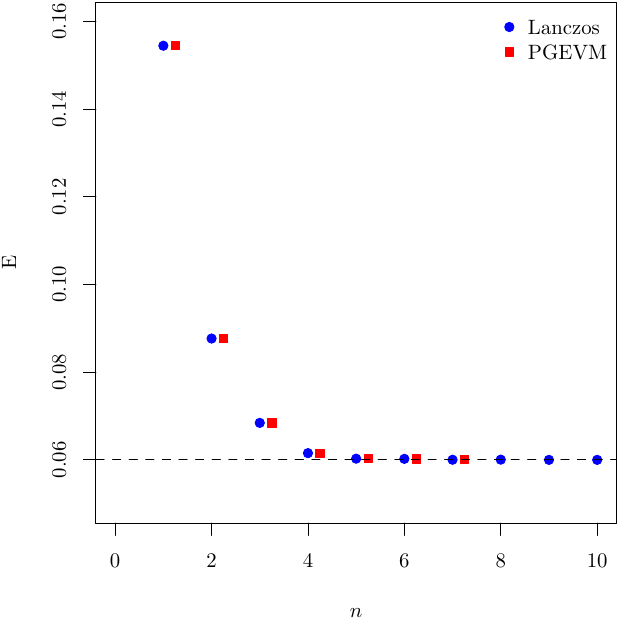}
  \includegraphics[width=0.45\textwidth, page=2]{plots/experiments.pdf}
  \caption{Comparison of PGEVM with $\delta t=1, \Delta=1, t_0=0$ and
    Lanczos for artificial data, see Eq.~\ref{eq:syn}. The PGEVM data is shifted slightly in
    $x$-direction for legibility. Left: convergence of the ground
    state energy level as a function of $n$. The exact value is
    indicated by the dashed line. Right: the
    difference to the exact ground state energy is plotted on a
    log-scale as a function of $n$.
    Empty symbols in the right panel indicate negative differences.}
  \label{fig:artificial1}
\end{figure*}

We first look at an artificially generated correlation function
\begin{equation}
  \label{eq:syn}
  C(t) = \sum_{i=0}^{N_s-1}\ e^{-E_i t}
\end{equation}
with $N_s=6$ and
\[
E_{0, 1, 2, 3, 4, 5} = \{0.06, 0.1, 0.13, 0.18, 0.22, 0.25\}\,.
\]
The results of PGEVM and oblique Lanczos are compared in
Fig.~\ref{fig:artificial1}. In the case of the PGEVM we use $\delta
t=1, \Delta=1, t_0=0$. In the left panel we plot the lowest energy level
as a function of $n$. Lanczos and PGEVM agree exactly up to round-off
until $n=6$, for 
which all $N_s$ states are resolved. Since we are using finite
precision arithmetics, we do not obtain the exact result, as visible
from the right panel, where the difference to the exact ground state
energy is plotted on a logarithmic scale, again as a function of $n$. 
PGEVM fails for $n>7$, because the inversion fails as the
corresponding Hankel matrix becomes singular. 

Notably, the Lanczos does not show this instability
and continues to converge until machine precision.
It is noteworthy that Lanczos does not converge monotonically any
longer for $n>8$, which can also be observed for the PGEVM once all
states are resolved.

This result confirms that Lanczos and PGEVM are equivalent.

\subsection{Nucleon Correlator}

\begin{figure*}
  \centering
  \includegraphics[width=0.45\textwidth, page=1]{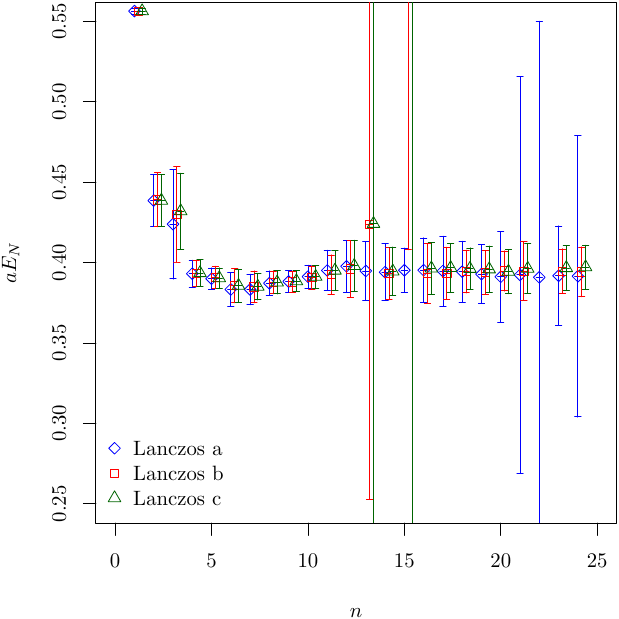}
  \includegraphics[width=0.45\textwidth, page=2]{plots/nucleon.pdf}
  \caption{Lanczos results for a nucleon two-point function. Left: comparison
    of Lanczos a,b,c results as a function of 
    $n$. Right: comparison of Lanczos a with the $\log$ effective mass
    definition.}
  \label{fig:nucleon1}
\end{figure*}

Let us now compare Lanczos and PGEVM for a nucleon two-point
function obtained from a real lattice QCD simulation. We have used
a $N_f = 2+1+1$ twisted mass lattice ensemble of size $64^3 \times 128$
with $a \approx 0.08$ fm and $m_\pi \approx 130$ MeV~\cite{Alexandrou:2018egz}.
The correlator $C(t) = \langle N(t) \bar{N}(0)\rangle$ is computed using
the local interpolating field $N(x) = \epsilon \, d(x) (u^T(x) C\gamma_5 d(x))$
and projected to zero momentum. We have used $200$ gauge configurations
with 16 sources each. First, in the left panel of 
Fig.~\ref{fig:nucleon1} we compare Lanczos a,b and c with
each other. All three perform very similarly with stable statistical
uncertainties also for $n$-values up to $20$, up to some outlier
$n$-values for Lanczos b and c. They
appear due to the bias correction applied in hindsight after the
bootstrap sampling with pivot element has been performed already. 
For $n>20$ the usage of the pivot mechanism helps in reducing the
uncertainties.

All three Lanczos variants share that the plateau is reached around
$n=4$. The energy estimate for all larger $n$-values is actually
compatible within errors with the estimate at $n=4$, with very little
variation. For Lanczos b and Lanczos c there is less increase in the
uncertainty, but some outliers appear where the eigenvalue
identification apparently failed.

In the right panel of the same figure we compare Lanczos a with the
standard $\log$ effective mass Eq.~\ref{eq:Meff}. The exponential
decrease in the StN ratio is clearly visible for the latter. The
plateau for $M_\mathrm{eff}$ is reached for $t=10$, using correlator
values as $t=10$ and $t=11$. The Lanczos method with $n=4$ uses all
correlator values up to $t=7$. Thus, both Lanczos and the
$\log$-effective mass reach the plateau once a certain $t$-value is
included in the analysis. 

\begin{figure}
  \centering
  \includegraphics[width=.45\textwidth]{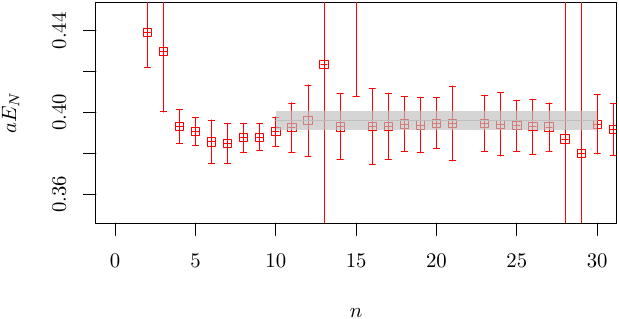}
  \caption{Lanczos b result for the nucleon as a function of $n$
    zoomed in on the $y$-axis. The solid line represents a possible
    plateau fit with error band.}
  \label{fig:nucleon2}
\end{figure}

When comparing the fluctuations of the expectation value with the
uncertainty estimate in Fig.~\ref{fig:nucleon1} one observes that
(apart from the outliers, which are due to eigenvalue
mis\-identification) errors are too large for the data points to be
independent. This is better visible in Fig.~\ref{fig:nucleon2}, where
we zoom in on the $y$-axis for Lanczos b and indicate a possible
plateau fit with the solid line and the error band.

\begin{figure*}
  \centering
  \includegraphics[width=0.45\textwidth, page=1]{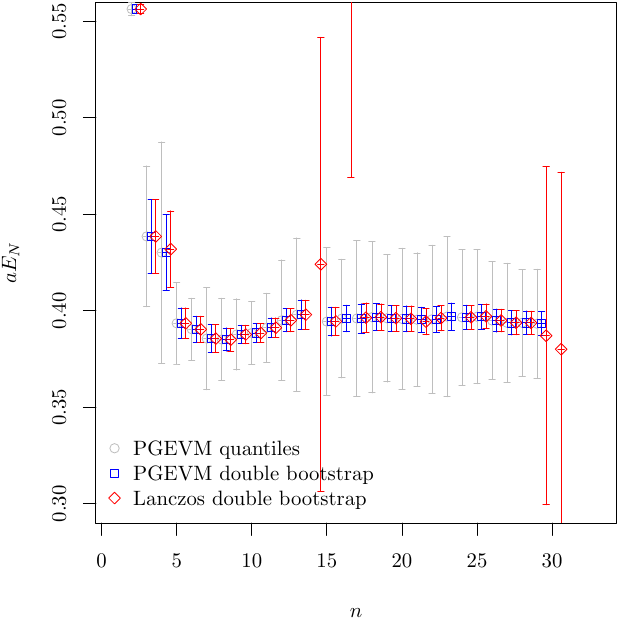}
  \includegraphics[width=0.45\textwidth, page=2]{plots/pgevm.pdf}
  \caption{Left:  Comparison of PGEVM with error estimate from
    quantiles, PGEVM with error estimate from double bootstrap and
    Lanczos with error estimate from double bootstrap.
    Right: Correlation of the estimates at different $n$-values as a function
    of $n'$.
  }
  \label{fig:nucleon3}
\end{figure*}

Thus, we next estimate the uncertainty using the double bootstrap as
discussed in the previous section. In the left panel of
Fig.~\ref{fig:nucleon3} we plot the ground state energy estimate as a
function of $n$ for PGEVM with quantiles as error estimate, and for
PGEVM and Lanczos with double bootstrap error estimate.

First of all, it becomes clear from this plot that PGEVM and Lanczos
with bias correction yield identical results also for noisy data,
apart from a few points where the eigenvalue identification
failed. Second, the double bootstrap uncertainty is significantly
smaller than the one from quantiles (and, therefore, also outlier
removal). Still, even if the error estimates are significantly
reduced, fluctuations are too small given the uncertainties for
independent data as reported in Ref.~\cite{Wagman:2024rid}.

While the estimate of the correlation of results at different
$n$-values appears difficult in the presence of (large and many)
outliers, the correlation can be computed more reliably on the double
bootstrap estimators. This correlation $\mathrm{Cor}(n, n')$ is plotted
in the right panel of Fig.~\ref{fig:nucleon3} for four values of $n$
as a function of $n'$. At $n=n'$ the correlation is identical to 1, as
it must be. For $n'<n$ we observe an almost linear increase towards
1. For $n'>n$, correlation drops quickly, but for large $n$ levels out
at a finite plateau value. This finite plateau value increases
significantly with $n$, reaching a value of $0.7$ for $n=20$.

A fully correlated, constant fit to the PGEVM (or Lanczos) double
bootstrap results in the range from $n=15$ to $n=29$ leads to $aE_N =
0.395(5)$ with a fairly large $p$-value of $0.999$, indicating again
the large correlation among the included data points. A fit to the
effective mass data from $t=10$ to $t=18$ results in $aE_N = 0.390(3)$
and a $p$-value of $0.64$, largely dominated by the effective mass
values at $t=10$ to $t=14$. Both fitted values are fully
compatible. The value quoted in Ref.~\cite{Alexandrou:2018egz} based on much higher
statistics is $aE_n = 0.3864(9)$.

In order to gain further insight into how the Lanczos method deals
with noisy correlator data at large $t$, we have carried out the
following experiment: we modify the two-point function by adding to
$C(t)$ a Gaussian distributed random shift with width equal to half of
the estimated error of $C(t)$. Likewise we modify the bootstrap samples
to preserve standard error and correlation and adapt the mean. 
Then we apply Lanczos to the modified correlator. 

\begin{figure*}
  \centering
  \includegraphics[width=0.45\textwidth, page=1]{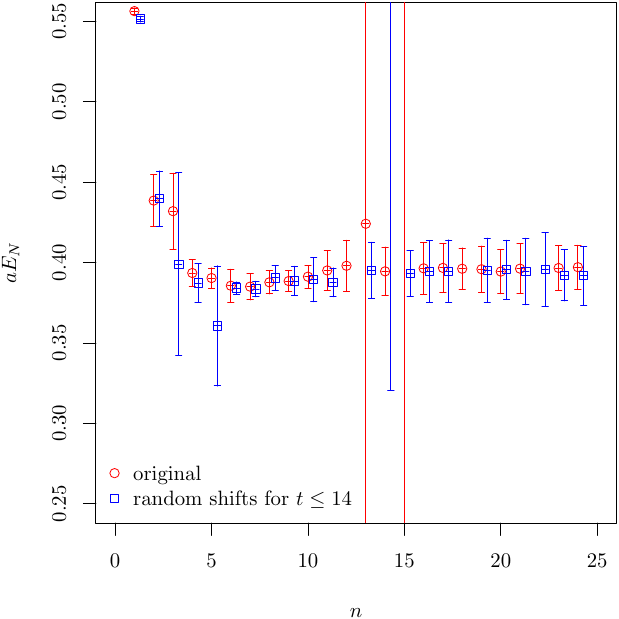}
  \includegraphics[width=0.45\textwidth, page=2]{plots/lanczos-shifts.pdf}
  \caption{Comparison of Lanczos b applied to the original correlator
    and to the ones which were modified for $t\leq 14$ (left) and for
    $t>14$ (right). The vertical dashed line in the right panel
    indicates the $n$-value from which on modified data enters the
    result.}
  \label{fig:nucleon_shifts}
\end{figure*}

The result is shown in Fig.~\ref{fig:nucleon_shifts}, in the left
panel the correlator is modified only for $t\leq 14$, in the right
panel for $t>14$. Random shifts for $t\leq 14$ are clearly visible in
the Lanczos result, while there is almost no change in the Lanczos
result on the correlator with modified data for $t>14$. Interestingly,
the random shifts for $t>14$ make the outlier disappear in the
original Lanczos result. One notes, however, that for both
modification scenarios the large $n$-result appears remarkably stable,
though still with very little variation of the expectation value
relative to the uncertainties.

There is another interesting observation to be made from
Fig.~\ref{fig:nucleon_shifts}: random variations of the correlator can
cure eigenvalue misidentification problems. For instance in the right
panel the results on the randomly shifted correlator for $t>14$ do not
show any misidentification issue anymore. This can be used to identify
misidentified eigenvalues: one can analyse one or two
correlators randomly modified for $t$-values larger than some
threshold as discussed above in addition to the original correlator
and check for stability of the result.

\subsection{Pion Correlator}

The pion correlator is, in contrast to the nucleon case, periodic in
time. For this case we use a pion correlator for the $N_f=2$ Wilson
twisted mass ensemble $B_1$ from Ref.~\cite{ETM:2009ztk} with $L=24$
and $T=48$. It has a pion mass value of about $300$~MeV at a lattice
spacing of $a=0.079$~fm. The correlator estimate is based on $316$ gauge
configurations. Note that this data set ships as a sample correlator
with the hadron package~\cite{hadron:2024}.

There are two main differences when compared to the nucleon: first, the
pion has no signal-to-noise problem; second, the corresponding
correlator is periodic. The latter has the consequence that both
\[
e^{-E_0 t}\,,\quad e^{E_0 t}
\]
appear in the spectral decomposition of the correlator with equal
amplitudes.

\begin{figure*}
  \centering
  \includegraphics[width=0.45\textwidth, page=1]{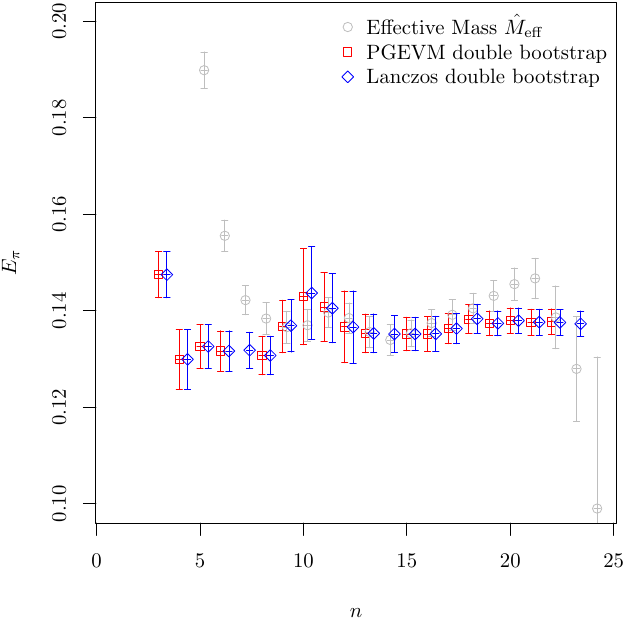}
  \includegraphics[width=0.45\textwidth, page=2]{plots/pion-pgevm.pdf}
  \caption{Pion results. Left: ground state pion energy $E_\pi$ as a
    function of $n$ for PGEVM and Lanczos, both with double
    bootstrap. In grey we we show for comparison
    $\hat{M}_\mathrm{eff}(t=n)$. Lanczos and effective mass data points are
    slightly shifted horizontally for legibility. Right: correlation
    estimates like the right panel of Fig.~\ref{fig:nucleon3}, but for the pion.}
  \label{fig:pion1}
\end{figure*}

The results of PGEVM and Lanczos, both with double bootstrap error
estimate, are compared in the left panel of Fig.~\ref{fig:pion1}. For
convenience we also plot the usual effective mass estimate
$\hat{M}_\mathrm{eff}(t=n)$. The general picture appears to be very
similar to the nucleon case with the exception of larger fluctuations
around $n=10$: this is an effect we attribute to the back propagating
state becoming relevant while the matrix size $n$ is not yet
sufficient to resolve all relevant states. This effect has been
observed already in Ref.~\cite{Fischer:2020bgv}.

Moreover, the correlation estimates for the
pion correlator are displayed in the right panel of
Fig.~\ref{fig:pion1}. Qualitatively, the behaviour appears similar to
the one of the nucleon correlator shown in the right panel of
Fig.~\ref{fig:nucleon3}. However, there are key differences. For $n=6$
correlation starts at zero for $n'=1$ and increases to 1 for $n'=6$,
as expected. Thereafter, the correlation does decrease again until
$n'=12$. For matrix sizes larger than $12$, the mirror part of the
correlator starts contributing, which will likely add little
information. This seems to be reflected by a (noisy) plateau in
$\mathrm{Cor}(6,n')$ for $n'>12$.

For $n=12$ the increase from $n'=1$ to $n'=12$ is present again, but
showing a sort of intermediate plateau between $n'=7$ and $n'=11$,
which is the region where the effective mass shows the plateau. For
$n'>12$ the correlation decreases again with maybe a plateau around
values of $0.4$. For $n=18$, finally, instead of the intermediate
plateau there is even a minimum in correlation around $n'=11$. Only
thereafter the correlation rises to $1$. For $n'>18$ the correlation
plateaus around a value of $0.8$.

Not surprisingly, a fully correlated constant fit to the PGEVM (or
Lanczos) results in a range from $n=7$ to $n=22$ basically reproduces
the result at $n=22$. The fit result reads $aE_\pi = 0.1378(24)$ with a 
reasonable $p$-value of $0.38$. This can be compared to the value
$aE_\pi =0.1362(7)$ quoted in Ref.~\cite{ETM:2009ztk} based on
larger statistics. Both values are fully compatible. If one were to
fit the effective mass shown in the left panel of
Fig.~\ref{fig:pion1}, it really depends on the fit range, but values
are also compatible, even though they tend to be higher than the PGEVM
result.

\section{Discussion}

The results presented in the last section clearly demonstrate that
(oblique) Lanczos and PGEVM yield identical results, not only for data
without noise, but also for noisy data, for which we looked at the
nucleon and the pion case separately. The nucleon exhibits the
signal-to-noise problem, while the pion is one of the few examples
that do not.

By using the double bootstrap estimate of the uncertainty of the
median estimator to the expectation value, the correlation of errors
of results at $n$ and $n'$ can be computed. For the nucleon case they
show clearly that the larger $n$ with $n'>n$ the results are
correlated and the correlation does plateau for $n'-n\gtrsim 3$. This
plateau value increases with increasing $n$-value.

This behaviour is entirely expected: increasing $n$ by 1 means two more
correlator values are included in the analysis. Thus, with increasing
$n$ the fraction of additional correlator values with new information
decreases like $1/n$. In addition, these correlator values have
increasingly large uncertainties. For instance, for $n=20$ there are
$40$ correlator values used, and $50$ for $n=25$. Thus, from $n=20$ to
$n'=25$ the fraction of additional correlator values is $1/5$ and we
would expect roughly $80\%$ correlation. We observe a little less. Due
to the increasing uncertainties in $C(t)$ itself this correlation is
not decreasing anymore and, hence, explaining the plateau.

One might still wonder why the exponentially increasing noise of the 
correlator values does not lead to an increase in the uncertainties of
Lanczos or PGEVM results at large $n$. We explain this as follows:
both Lanczos and PGEVM can also work on data with imaginary
eigenvalues. And it appears that most of the noise is mapped to those
imaginary eigenvalues once the physical eigenvalues are determined
sufficiently from correlator values at smaller $t$.

One point worth mentioning here is that eigenvalue identification might
still fail despite the methods applied here. We observed in our
experiments that random variations of the data can be a way to cure
this problem: if after a variation of the correlator within its
uncertainties one of the eigenvalues is not found again, it is with
high probability a spurious eigenvalue. This could be an alternative
to the method described in Ref.~\cite{Wagman:2024rid} where the first
row and column of $T_n$ are removed, eigenvalues are recalculated, and
spurious eigenvalues identified with the same reasoning.

For the pion, on the other hand, without the signal-to-noise problem
there appears to be a gain in information up to half of the time
extent, after which the correlator is mirrored leading to little
additional information.

Finally, we observed for both the pion and the nucleon correlator the
behaviour that the PGEVM and Lanczos results show a local minimum at
early $n$-values, see Fig.~\ref{fig:nucleon3} and
Fig.~\ref{fig:pion1}. Currently, we cannot explain this behaviour.

\section{Summary}

In this paper we have discussed the equivalence of the Prony
GEVM~\cite{Fischer:2020bgv} and Lanczos~\cite{Wagman:2024rid}
methods. They are exactly equivalent not only on the 
analytical level, but also in practice even for noisy data. In particular,
both methods converge equally fast to the lattice energy levels one is
interested in. We have also seen that double bootstrap is absolutely
essential for a robust uncertainty estimate. With double bootstrap and
the bootstrap median as the estimator for the expectation value there
is also no empirical outlier removal procedure required.

Unfortunately, we conclude from our analysis of the error
distributions that the Lanczos method does not solve the
signal-to-noise problem. With uncertainties estimated from double
bootstrap the correlation of results including correlators at larger
and larger $t$-values becomes visible. 

One clear advantage PGEVM and Lanczos do have when compared
to the standard effective mass is that the result at a large enough
$n$-value with its error gives a fit range independent estimate of the
energy value. Likely, a model averaging procedure is obsolete in this
case. We also observe that both PGEVM and Lanczos provide a lower
estimate of the ground state energy level than a plateau fit to the
effective mass on the same data. Moreover, this lower value is closer to
the estimate obtained with the effective mass approach based on
significantly higher statistics. 

One should also note that the PGEVM at fixed $n$ scales for large $n$
like $\mathcal{O}(n^3)$, while the Lanczos method scales only like $\mathcal{O}(n^2)$. For
ensembles with large time extent this is a clear advantage of the
Lanczos method. However, PGEVM is more general in principle by adding
more degrees of freedom to the algorithm.

Finally, we would like to point out that the block PGEVM introduced in
Ref.~\cite{Fischer:2020bgv} should be equivalent to a block Lanczos
approach based on the same arguments as discussed in this paper.

~\\\section*{Acknowledgments}
We thank Daniel Hackett for very useful discussions.
This work was supported by the
Deut\-sche Forschungsgemeinschaft (DFG, German Research Foundation) as
part of the CRC 1639 NuMeriQS – project no. 511713970.
The open source software packages R~\cite{R:2019} and hadron~\cite{hadron:2024}
have been used.

\printbibliography

\end{document}